\documentstyle[12pt]{article}
\newcommand{\be}{\begin{equation}}
\newcommand{\ee}{\end{equation}}
\newcommand{\no}{\noindent}
\begin{document}
\begin{flushright}
{\bf hep-th/0008049}\\
{\bf UCVFC-DF/18-99}
\end{flushright}
\vskip 1cm
\begin{center}
{\LARGE Maxwell Chern Simons Theory in a Geometric Representation}

\vskip 12pt

{\Large Lorenzo Leal and Oswaldo Zapata}

{\it Departamento de F\'{\i}sica, Facultad de Ciencias, \\
Universidad Central de Venezuela, AP 47270,\\
Caracas 1041-A, Venezuela}\\{\em Email: {\tt lleal@tierra.ciens.ucv.ve}}\\   
\end{center}
\abstract{
We quantize the Maxwell Chern Simons theory in a geometric representation
that generalizes the Abelian Loop Representation of Maxwell theory. We find
that in the physical sector, the model can be seen as the theory of a massles
scalar field with a topological interaction that enforces  the wave functional
to be multivalued. This feature allows to relate the Maxwell Chern Simons
theory with the quantum mechanics of particles interacting through a Chern
Simons field.}

\newpage

\section{Introduction}

The Maxwell Chern Simons theory (M.C.S.T.) \cite{deser} presents the interesting
property of being a massive theory, althought being gauge invariant. The mass
is provided by the topological Chern-Simons (C.S.) term, which, in turn, has
been widely considered both in the abelian and non abelian cases as a pure
gauge theory \cite{deser, hagen, witten, dunne, labastida}.

\noindent The purpose of this letter is to study the geometric representation
appropriate for the abelian M.C.S.T. theory, within the spirit of the Loop
Representation of Maxwell Theory \cite{gaetano, gambinitrias, librogambini}.
Our motivation is mainly to obtain a further insight into the loop and path
representations, that could be useful for later developments
in more realistic theories, such as Quantum Gravity in the Ashthekar
formulation \cite{ashtekar, rovellismolin, rodolfo}. Within this program, the
path representation of the Proca-Stueckelberg theory was studied recently
\cite{camacaro}. Also, the Maxwell field coupled to point particles has
been quantized in a geometric representation \cite{ryan}. A common feature of
these  models, which is also shared by the free Maxwell theory, is that the
introduction of loops or open paths (depending on the case) automatically
solves the Gauss constraint. As we shall see, this is not the case for the
M.C.S.T.. Instead, the Gauss constraint further restricts the path space,
leaving the boundary of the paths as the relevant geometric structures, except
by the fact that the theory is sensible to the number of times hat the paths
wind around their own boundaries. This feature lead to deal with multivalued
wave functionals. A similar result was obtained for the Chern Simons field
coupled to a scalar field several years ago \cite{loli}.

\noindent The multivaluedness of wave functions due to topological interactions 
is the hallmark of anyonic behavior within the context of quantum
mechanics \cite{leinaas, wilczek, wu, arovas, iengo, jackiwernesto, jackiwpi}.
Hence, one could interpret the M.C.S.T. as one of point particles,
lying at the boundaries of the paths, and obeying
fractional statistics. The statistical parameter results to be related to
the mass of the model. Indeed, the mass term can be gauged away by the
singular gauge transformation that maps the ordinary wave function into the
multivalued one. At last, the geometric approach allows to display the following
equivalence: the M.C.S.T. may be mapped into a massles scalar field theory
with fractional statistics.

\noindent The organization of the paper is as follows. In section 2 we recall
some basic results about the M.C.S.T. and its canonical quantization. In
section 3  we review the abelian path space, and study the path representation
of the quantum M.C.S.T., paying special interest to the geometric
resolution of the Gauss constraint. Section 4 is devoted to explore the
relation between the M.C.S.T., the massles scalar field theory and the quantum
mechanics of non relativistic particles with C.S. interaction. A short
discussion is presented in section 5.

\section{The Model}

The M.C.S. lagrange density is given by \cite{deser}

\begin{equation}
 {\mathcal L} =-\frac{1}{4}F_{\alpha \beta} F^{\alpha \beta} +
\frac{k}{4\pi} \epsilon
^{\alpha \beta \gamma} \partial_\alpha A_\beta A_\gamma
\label{eq:lagrangeano}
\ee                              
                                                       
\noindent where $F_{\alpha \beta}=  \partial_\alpha A_\beta -
\partial_\beta A_\alpha$. We take $g_{\mu \nu}= diag(1,-1,-1)$.
 The equation of motion 

\be
\partial_{\alpha}F^{\alpha \gamma} + \frac {k}{2\pi} \epsilon^{\alpha
\beta \gamma} \partial_{\alpha} A_{\beta} = 0
\label{eq:ec de mov} 
\ee

\noindent leads to

\begin{equation}
(\Box +(\frac{k}{2\pi})^2)F_{\alpha \beta} = 0
\label{eq:dalambert}
\end{equation}

\noindent which shows that the M.C.S. gauge field is massive ($k$ has units of
mass, as can be readily seen from eq.(\ref{eq:ec de mov}). Moreover, it can
be shown that the theory posseses a single excitation with mass 
$\frac{\mid k\mid}{2\pi}$ and spin $1$ ($k>0$) or $-1$ ($k<0$) \cite{deser}.

\noindent The canonical quantization {\em a la Dirac} yields the following
results. In the Weyl gauge ($A_0 = 0$) there is a first class constraint generating
 the time independent gauge transformations

\begin{equation}
\partial_{i} \Pi^{i}({\bf x}) +\frac{k}{4\pi} B({\bf x}) = 0
\label{eq:ligadura}
\end{equation}

\noindent with $B = \epsilon^{ij}\partial_{i}A_{j}$, and $\Pi^i$ being the
 canonical momentum satisfying

\begin{equation}
[A_{i}({\bf x}),\Pi^{j}({\bf y})] = i \delta_i^j \delta^2 ({\bf x}-{\bf y})
\label{eq:conmutador}
\end{equation}

\noindent The remaining canonical commutators vanish.

\noindent Unlike the pure Maxwell case, the momentum $\Pi^i$ does not coincide
 with the electric field. Indeed

\begin{equation}
E^i = F^{i0} = \Pi^i - \frac{k}{4\pi} \epsilon^{ij}A_{j}
\end{equation}

\noindent It is easily verified that both $E^i$ and $B$ are gauge invariant
quantities, in contrast with $ \Pi^i$ and $A_{i}$. The algebra of the
 fundamental observables results to be

\begin{equation}
[E^{i}({\bf x}),B({\bf y})] = -i \epsilon^{ij}\partial_j \delta^2
 ({\bf x}-{\bf y})
\end{equation}

\be
[E^{i}({\bf x}),E^{j}({\bf y})] =- i \frac{k}{2\pi} \epsilon^{ij}
 \delta^2 ({\bf x}-{\bf y})
\ee

\be
[B({\bf x}) , B({\bf y})] = 0
\ee

\noindent The Hamiltonian of the theory, on the physical sector, is given by

\be
H = \int d^2x \frac{1}{2}[(E^i)^2 + B^2]
\label{eq:hamiltoniano}
\ee

\noindent which, together with the conserved momentum

\be
P^i = -\int d^2x\, \epsilon^{ij} E^j B
\ee

\noindent the angular momentum

\be
J = \int d^2x\, x^i E^i B
\ee

\no and the generator of Lorentz boosts

\be
M^{i0} = \frac{1}{2} \int d^2x [(E^i)^2 + B^2] - tP^i
\ee

\no provide a representation of the Poincare Algebra in 2+1 dimensions
\cite{deser}.

\section{Path Space Representation}

Now we focus on the geometric representation appropriate to the M.C.S.T..
To this end, we recall some basic facts about the
path-representation \cite{camacaro}. Given a curve $\gamma$ in $R^n$, we define
its form factor

\be
T^i (x,\gamma) = \int_{\gamma} \delta^n(x-y) dy^i
\label{eq:factorforma} 
\ee

\no which is independent of the parametrization chosen. It should be said
that $\gamma$ could consists on several disjoint pieces, some of which could
be closed. Expression (\ref{eq:factorforma}) allows to group the curves in
equivalence classes: two curves $\gamma$ and $\gamma'$ are said to be
equivalent if $T^i(\gamma)=T^i(\gamma')$ . It is a simple matter to show that
this is indeed an equivalence relation. The equivalence classes of curves
$[\gamma]$ are denominated paths. From now on, we shall not make distinction
between a path and any of its representatives.

\no The usual composition of curves can be lifted to a group product among
paths as follows. Given two curves $\gamma_1$,$\gamma_2$, the form factor of
their composition $T^i(\gamma_1.\gamma_2)$ does not depend on the
representatives, i.e.: $T^i(\gamma_1'.\gamma_2')= T^i(\gamma_1.\gamma_2)$,
whenever $\gamma_1'\sim \gamma_1$ and  $\gamma_2'\sim \gamma_2$. Hence,
we define the product of two paths as the class to which the composition of
their representative curves belong.
Furthermore, the equivalence class of the opposite curve $-\gamma$ plays the
role of the inverse path, while the equivalence class of the null curve
amounts to the identity. The group so defined is Abelian, as may be readily
seen.

\no As in the study of the Proca field \cite{camacaro}, we shall use the
path derivative $\Delta_i(x)$, which measures the change of a path-dependent
functional $\Psi(\gamma)$ when a small open path $\delta\gamma_{x}^{x+h}$
starting at $x$ and ending at $x+h$ ($h\rightarrow 0$) is attached to $\gamma$ :

\be
\Psi(\delta\gamma.\gamma) \equiv (1 + h^i \Delta_i(x)) \Psi(\gamma)
\label{eq:derivada camino}
\ee

\no Equation (\ref{eq:derivada camino}), defining $\Delta_i(x)$, must be
thought to hold up to first order in $h^i$. The path derivative is related to
the abelian loop derivative $\Delta_{ij}(x)$ of Gambini-Trias
\cite{gambinitrias} by

\be
\Delta_{ij}(x)  =  \frac{\partial}{\partial x^i}\Delta_j(x) -
                     \frac{\partial}{\partial x^j}\Delta_i(x)  
\ee

\no This last object, also known as the area derivative, serves to compute
how a path (or loop) dependent functional changes when a small plaquette is
attached to it at the point $x$.

\no Using definition (\ref{eq:derivada camino}) it is a trivial matter to show
that (we return to 2+1 dimensions)

\be
\Delta_i({\bf x}) T^j({\bf x}',\gamma) = \delta_i^j \delta^2({\bf x}-{\bf x}')
\ee

\no hence, the canonical algebra (\ref{eq:conmutador}) is fulfilled if we set

\be
A_i({\bf x}) \longrightarrow \frac{i}{e} \Delta_i({\bf x})
\label{eq:realiza1}
\ee

\be
\Pi^i({\bf x}) \longrightarrow eT^i({\bf x},\gamma)
\label{eq:realiza2}
\ee

\no which constitutes a realization of the canonical operators onto path
dependent wave functionals $\Psi(\gamma)$. In equations (\ref{eq:realiza1})
and (\ref{eq:realiza2}) , the constant $e$ with units of $[mass]^{\frac{1}{2}}$
was introduced to properly adjust the dimensions.

\no To write down the constraint eq. (\ref{eq:ligadura}) in the
path-representation, we need to calculate:

\begin{eqnarray}
\frac{\partial}{\partial x^i} T^i({\bf x},\gamma)
             &  = & -\sum_s(\delta^2 ({\bf x}-{\bf \beta_s})
                 -\delta^2 ({\bf x}-{\bf \alpha_s}) )\nonumber\\
             & \equiv & - \rho(x,\gamma)
\end{eqnarray}

\no where $\beta_s$ ($\alpha_s$) is the ending (starting) point of the piece s
which contributes to the whole path $\gamma$ (remember that $\gamma$  may
consist of several disjoint pieces). Thus,  $ \rho({\bf x},\gamma)$ can be
thought as the ``form factor'' of the boundary of the path. The first class
constraint of the theory eq.(\ref{eq:ligadura}) demands that the physical
(i.e., gauge invariant) wave functionals obey

\be
\left(-\rho({\bf x},\gamma) + \frac{ik}{8\pi e^2}
\epsilon^{ij}\Delta_{ij}({\bf x})\right) \Psi(\gamma)= 0
\label{eq:ligacaminos}
\ee

\no It is worth mentioning a mayor difference between the geometric
representation of the M.C.S. and the pure Maxwell or Proca theories. In the
Maxwell case \cite{gaetano} , the introduction of loop-dependent functionals
automatically solves the gauge constraint. Similarly, the use of path-dependent
wave functionals fulfiles gauge invariance in the Proca-Stueckelberg theory
\cite{camacaro}. However, this is not the case with the M.C.S.T.. Further
restrictions on the path dependence of $\Psi(\gamma)$  which are to be dictated
by the constraint eq.(\ref{eq:ligacaminos}) remain to be considered. To this
end, we set, without lose of generality

\be
\Psi(\gamma) = exp(i\chi(\gamma)) \Phi(\gamma)
\ee

\no and ask $\chi(\gamma)$ to obey

\be
\epsilon^{ij}\Delta_{ij}({\bf x})\chi(\gamma)= -\frac{8\pi e^2}{k}\rho({\bf x},\gamma)
\label{eq:inhomogenea}
\ee

\no then eq.(\ref{eq:ligacaminos}) reduces to

\be
\epsilon^{ij}\Delta_{ij}({\bf x})\Phi(\gamma)= 0
\label{eq:homogenea}
\ee

\no Equation (\ref{eq:inhomogenea}) is solved by

\begin{eqnarray} 
\chi(\gamma) & = &\frac{2e^2}{k}\int d^2x \int dx'^2 \partial_i\partial_l
                 ln|{\bf x}-{\bf x}'|\epsilon^{lk}T^i({\bf x}',\gamma)T^k
                ({\bf x},\gamma)\nonumber\\
             & = & -\frac{2e^2}{k} \sum_s \int_{\gamma}dx^k \epsilon^{lk}
                     \left[\frac{(x-\beta_s)^l}{|{\bf x}-{\bf \beta_s}|^2} -
                     \frac{(x-\alpha_s)^l}{|{\bf x}-{\bf \alpha_s}|^2}\right]
\label{eq:chi}
\end{eqnarray}

\no as a careful application of the area derivative shows.

\no Since

\be
\theta = \int_{\gamma} dx'^k \epsilon^{lk} \frac{(x'-x)^l}{|{\bf x}'-{\bf x}|^2}
\ee

\no is the angle subtended by the path  $\gamma$ from the point ${\bf x}$, we see
that eq. (\ref{eq:chi}) yields

\be
\chi(\gamma) = - \frac{2e^2}{k} \Delta\Theta
\ee

\no where $\Delta\Theta$ is equal to the sum of the angles subtended by the
pieces of the path from their final points $\beta_s$ , minus the angles
subtended by the same pieces measured from their starting points $\alpha_s$.
Hence, we see that $\chi(\gamma)$  depends on $\gamma$ through their boundary,
and through the way that the diverse pieces of $\gamma$ wind around these
boundary points $\alpha_s\,'s$ and $\beta_s\,'s$.

\no Equation (\ref{eq:homogenea}), on the other hand,  states that
$\Phi(\gamma)$ is insensitive to the addition of closed paths,
i.e., $\Phi(C.\gamma) = \Phi(\gamma)$ , where $C$ is a loop. Thus,
$\Phi(\gamma)$ only depends on the boundary of the path:

\be
\Phi(\gamma) =  \Phi(\alpha_s ;\beta_s)
\ee

\no Summarizing, we have that on the physical sector

\be
\Psi(\gamma_{\alpha}^{\beta})_{Physical}  = exp\, (-i\frac{2e^2}{k}\Delta
\Theta)\,
\Phi(\alpha_s ;\beta_s)
\label{eq:soluliga}
\ee

\no with $\Phi(\alpha_s ;\beta_s)$ an arbitrary functional of the boundary
of the path.

\no Expression (\ref{eq:soluliga}) is then the solution to the gauge constraint
in path space eq.({\ref{eq:ligacaminos}). We see that although the introduction
of paths does not solve automatically the constraint, it does allow to
characterize the physical sector in a geometrically appealing form. To write
down the physical observables of the theory in the path space representation,
one needs to know how the gauge invariant operators $B$ and $E^i$ act onto the
physical sector of the Hilbert space. One has, after some calculations

\begin{eqnarray}
& &  E^i({\bf x})\,exp(i\chi(\gamma))\, \Phi(\alpha_s ;\beta_s)
     = exp(i\chi(\gamma))\,\times\nonumber\\
& &   \qquad \qquad  \times \,\left[-\frac{e}{\pi} \sum_s 
                     \left(\frac{(x-\beta_s)^i}{({\bf x}-{\bf \beta_s}|^2} -         
                     \frac{(x-\alpha_s)^i}{|{\bf x}-{\bf \alpha_s}|^2}\right)
                    -\frac{ik}{4\pi e} \epsilon^{ij}\Delta_j({\bf x})\right]
                     \Phi(\alpha_s ;\beta_s)
\label{eq:E} 
\end{eqnarray}

\no and

\be
B({\bf x})\,exp(i\chi(\gamma))\, \Phi(\alpha_s ;\beta_s)\, = \frac{4\pi e}{k}
 exp(i\chi(\gamma))\, \rho({\bf x},(\alpha;\beta)) \Phi(\alpha_s ;\beta_s)
\label{eq:B}   
\ee
where we have set $\rho({\bf x},\gamma) = \rho({\bf x},(\alpha;\beta))$ to
stress the fact that $\rho$ depends on $\gamma$ just through its boundary, the
set of starting points $\alpha_s$ and ending points $\beta_s$.

\no We thus see that the physical sector is invariant under the action of
both $B$ and $E^i$, as expected. It should be remarked that the path derivative
$\Delta_i({\bf x})$ acting on $\Phi(\alpha_s ;\beta_s)$  is a well defined
object, since a boundary dependent function $\Phi(\partial\gamma)$ is a special
kind of a path-dependent one. From eqs. (\ref{eq:soluliga}-\ref{eq:B}) we also
see that there is a simple unitary transformation which allows to eliminate the
path dependent phase , namely:

\begin{eqnarray}
\Psi(\gamma)_{Physical} & \rightarrow & \tilde{\Psi}(\gamma) =
           exp(-i\chi(\gamma))\, \Psi(\gamma)_{Physical}= 
           \Phi(\alpha,\beta)\nonumber\\
A_{Physical} & \rightarrow & \tilde{A} =  exp(-i\chi(\gamma))
A_{Physical} \,exp(i\chi(\gamma))
\label{eq:unitaria}
\end{eqnarray}

\no where $A_{Physical}$ is any gauge invariant operator of the theory. After
the unitary transformation is performed, what is left is a dependence on the
set of ``signed'' points $\alpha_{s}$ and $\beta_{s}$, corresponding to the
boundary of the missed path. It can be shown that these sets of signed points
inherit a group structure due to the paths where they come from. I fact,
when a starting  and ending points meet at the same place, they anihilate.
Therefore we shall refer to them as ``points'' and ``anti-points'' respectivelly.
It is worth saying that there is a non-abelian version of this group of points,
which encodes the kynematics of the Principal Chiral Fields, and that will be
discussed elsewhere.

\no From eqs. (\ref{eq:hamiltoniano}, \ref{eq:E}, \ref{eq:B}), and taking 
into account the unitary transformation eq.(\ref{eq:unitaria}), we can
write down the Schr\"{o}dinger equation in the geometric representation

\newpage

\begin{eqnarray}
& &  i\frac{\partial}{\partial t}\, \Phi((\alpha_s ;\beta_s),t)
     = \nonumber\\
& &   \qquad \qquad  = \int dx^2 \left[ \left[
-\frac{e}{\pi} \sum_s
                     \left(\frac{(x-\beta_s)^i}{|{\bf x}-{\bf \beta_s}|^2} -
                     \frac{(x-\alpha_s)^i}{|{\bf x}-{\bf \alpha_s}|^2}\right)
                    -\frac{ik}{4\pi e}
  \epsilon^{ij}\Delta_j(x) \right]^2 \right. \nonumber\\
& &   \qquad \qquad  \left. +{\left(\frac{4\pi e}{k}\right)}^2 \rho^2
(x,(\alpha;\beta))
        \right]   \Phi((\alpha_s ;\beta_s),t)
\label{eq:schrodinger}
\end{eqnarray}

\no In a similar way, the conserved momentum $P^i$, angular momentum $J$ and
the boosts generators $M^{0i}$ can be realized in the geometric representation.
It may be seen that the operators $P^i$ and $J$ act by translating and rotating
the argument of the wave functional $\Phi(\alpha;\beta)$ ; for instance

\be
(1+u^i P_i)\, \Phi(\alpha;\beta) = \Phi({\bf \alpha +u} ;\, {\bf \beta +u})
\ee

\no with ${\bf u}$ being an infinitesimal constant spatial vector. It must be
said that both $P^i$ and $J$ , inasmuch $H$ should be properly regularized,
since they involve ill defined products of distributions (needless to say that
this feature is not a consequence of the geometric representation).

\section{Relation with the Massles Scalar Field and Nonrelativistic Anyons}

\no The Schr\"{o}dinger equation (\ref{eq:schrodinger}) resembles the wave
equation of a collection of point particles interacting through a Chern-Simons
term \cite{arovas, jackiwernesto, jackiwpi}, in the sense that there appears a
``covariant derivative''

\be
-i D_l({\bf x}) \equiv -i \Delta_l ({\bf x}) - \frac{4e^2}{k}
\sum_s \,\epsilon_{il}
                \left(\frac{(x-\beta_s)^i}{|{\bf x}-{\bf \beta_s}|^2} -
                \frac{(x-\alpha_s)^i}{|{\bf x}-{\bf \alpha_s}|^2}\right)
\label{eq:derivacovar}
\ee

\no which comprises, besides the path-derivative $\Delta_j ({\bf x})$ , a term
of statistical interaction among the points $\alpha$ and antipoints $\beta$ ,
which should play the role of the particles. This observation can be made more
precise as follows. Let us consider the {\em singular} gauge transformation

\be
\Phi(\alpha;\beta) \rightarrow \bar{\Phi} (\alpha;\beta) \equiv
 exp(i\Lambda(\alpha;\beta)) \Phi(\alpha;\beta)
\label{eq:gaugesingular}
\ee

\no with

\begin{eqnarray}
 \Lambda(\alpha;\beta)) & = & - \frac{2e^2}{k} \int d^2 x \int d^2 y \,\,
        \rho({\bf x},(\alpha;\beta)) \,\theta({\bf x}-{\bf y})\,
        \rho({\bf y},(\alpha;\beta))\nonumber\\
                        & = & - \frac{2e^2}{k} \sum_s \sum_{s'}\,
        [ \theta ({\bf \beta_s} -\beta_{s'})
        -\theta (\beta_s -\alpha_{s'})+\theta (\alpha_s -\alpha_{s'})
        -\theta (\alpha_s -\beta_{s'})]\nonumber\\
\label{eq:definelambda}
\end{eqnarray}

\no $\theta({\bf x})$ being the angle that the vector ${\bf x}$ makes with the positive
$x-axis$ . With the aid of the expressions

\be
\Delta_i({\bf z}) \rho({\bf x},(\alpha;\beta)) = \frac{\partial}{\partial z^i}
 \delta^2 ({\bf z}-{\bf x})
\ee

\no and 

\be
\frac{\partial}{\partial x^l} \theta({\bf x}) = -\epsilon_{lk} \,
\frac{x^k}{|{\bf x}|^2}
\ee

\no it can be seen that

\be
-iD_j({\bf x}) \,\Phi(\alpha;\beta) = -i\, exp\,(i\Lambda(\alpha;\beta))\,
 \Delta_j({\bf x}) \,\bar{\Phi}(\alpha;\beta)
\ee

\no Thus, in the covariant derivative and in the Schr\"{o}dinger equation, the
interaction may be removed at the expense of dealing with redefined wave
functionals $\bar{\Phi}$, which result to be multivalued. In fact, the
Schr\"{o}dinger equation for that multivalued wave functional may be writen as

\begin{eqnarray}
& &  i\frac{\partial}{\partial t}\,\bar{ \Phi}((\alpha_s ;\beta_s),t)
     = \nonumber\\
& &   \qquad \qquad  = \int dx^2 \left[
                    -(\frac{k}{4\pi e})^2
  (\Delta_j({\bf x}))^2  +{(\frac{4\pi e}{k})}^2 \rho^2 ({\bf x},(\alpha;\beta))
        \right] \bar{\Phi}((\alpha_s ;\beta_s),t)
\label{eq:schrodingerm}
\end{eqnarray}

\no which corresponds to the Schr\"{o}dinger equation for the massles scalar
field theory, with lagrange density

\be
{\cal L}_{\phi} = \frac{1}{2} \,\partial_\mu \phi \,\partial^{\mu} \phi
\ee

\no in a geometric representation, as we briefly discuss. The associated
hamiltonian is

\be
H_{\phi} = \int d^2x \, \frac{1}{2}\,({\Pi}^2 + \partial_i \phi\,
 \partial_i \phi)
\ee

\no with $\Pi$ being the canonical momentum

\be
[\phi({\bf x}) , \Pi({\bf y})] = i\delta^2 ({\bf x}-{\bf y})
\label{eq:conmutfi}
\ee

\no If we prescribe the realization

\be
 \partial_i \phi({\bf x}) \rightarrow -i \Delta_i({\bf x})
\ee

\be
\Pi ({\bf x}) \rightarrow \rho({\bf x},(\alpha;\beta)) 
\ee

\no onto wave functionals $\Phi_{\phi} (\alpha;\beta) $, the commutator
(\ref{eq:conmutfi}) is verified, as well as

\be
[\partial_i \phi({\bf x}) \, , \partial_j \phi({\bf y})]\, = \, 
[\Pi({\bf x}), \Pi({\bf y})] \,= \,0
\ee

\no while the corresponding Schr\"{o}dinger equation reads

\begin{eqnarray}
& &  i\frac{\partial}{\partial t}\, \Phi_{\phi}((\alpha_s ;\beta_s),t)
     = \nonumber\\
& &   \qquad \qquad  = \int d^2x \left[
                    -(\frac{k}{4\pi e})^2
  \Delta_j({\bf x})^2  +{(\frac{4\pi e}{k})}^2 \rho^2 ({\bf x},(\alpha;\beta))
        \right]   \Phi_{\phi}((\alpha_s ;\beta_s),t)
\label{eq:schrodingerscalar}
\end{eqnarray}

\no which coincides with eq.(\ref{eq:schrodingerm}), as claimed, except by the
fact that here the wave functional $\Phi_\phi $ is single valued. It should be
observed that there is no need to realize $\phi$, since only
its derivative $\partial_i \phi$ appears in the expressions for the
observables of the theory. This reflects the invariance of the model
under the shift $\phi \, \rightarrow \phi + constant $.

\no The fact that the path dependence is only manifested through the boundary 
$(\alpha ; \beta)$ of the path $\gamma$ , evidences that, indeed, there is a
simpler geometry underlying both the M.C.S. and the massles scalar field
theories: the appropriate geometric representation is one of sets of
``points'' and ``anti-points'' (see comment after eq.(\ref{eq:gaugesingular})).
This signed point group is the first member in a list of geometric structures
related to gauge theories of $p-forms$ , to which paths and loops (for the case
$p = 1$) belong.

\no We are ready to compare the M.C.S. Schr\"{o}dinger equation in
path-space, eq.(\ref{eq:schrodingerm}) and its multivalued wavefunction
$\bar \Phi$ , eq.(\ref{eq:gaugesingular}), with what results from the
quantization of a collection of $N$ non-relativistic particles interacting
through a C.S. term \cite{arovas} \cite{jackiwernesto}, \cite{jackiwpi}. The
corresponding Schr\"{o}dinger equation may be writen as

\be
i \partial_t \Psi ({\bf r_1},...{\bf r_N},t) = \sum \limits_{p=1}^{N} 
-\frac{1}{2m_p}\,({\bf \bigtriangledown_p} - i e_p {\bf a_p})^2
\,\Psi ({\bf r_1},...{\bf r_N},t)
\label{eq:particles}
\ee 

\no with

\be
{\bf a_p} =  \frac{1}{k}\, {\bf \bigtriangledown_p}
 \sum \limits_{p \not= q}^{N} e_q \,\theta_{pq}  
\ee

\no while $\theta_{pq}$ is the angle that the vector ${\bf x_p}-{\bf x_q}$
makes with the $x-axis$.

\no Equation (\ref{eq:particles}) may be written in the form

\be
i \partial_t \Psi_0 ({\bf r_1},...{\bf r_N},t) = \sum \limits_{p=1}^{N}  
-\frac{1}{2m_p}\,({\bf \bigtriangledown_p})^2
\,\Psi_0 ({\bf r_1},...{\bf r_N},t)
\ee

\no with

\be
\Psi_0 = exp\, (-i \sum \limits_{p<q}\, \frac{e_p e_q}{k}\, \Theta_{pq})\,
\Psi
\label{eq:define psi0}
\ee

\no The multivalued function $\Psi_0$, which converts the the multiparticle
Schr\"{o}dinger equation into a ``free'' equation, presents remarkable
coincidences with the functional $\bar {\Phi} (\alpha; \beta)$ (eqs.
(\ref{eq:gaugesingular} , \ref{eq:definelambda})). In fact, since\,
$\theta (\beta_s -\beta_{s'})\,=\,\theta (\beta_{s'} -\beta_s) \pm 2\pi$,\,
equation (\ref{eq:definelambda}) can be writen as:

\begin{eqnarray}
 \Lambda(\alpha;\beta)& = & - \frac{e^2}{k} \sum \limits_{s'<s}\,
        [ \theta ({\bf \beta_s} -\beta_{s'})
        -\theta (\beta_s -\alpha_{s'})+\theta (\alpha_s -\alpha_{s'})
        -\theta (\alpha_s -\beta_{s'})] + const. \nonumber\\
\label{eq:lambda}
\end{eqnarray}

\no In writing eq. (\ref{eq:lambda}) we have omited the
undetermined ``self-interaction'' terms of\, the \,type 
$\theta (\beta_s-\beta_s)= \theta (0)$. If the charges of the particles in
the CS-point-particles theory are restricted by $|e_p|= e$, we see that the
phase in eq. (\ref{eq:define psi0}) coincides with $ \Lambda(\alpha;\beta)$.
In both cases, exchange of two ``particles'' makes the wave function to pick up
a phase factor which is a multiple of $exp (\pm i\frac{e^2 \pi}{k})$, depending
on the route followed to exchange the ``points'' and ``anti-points'' (or the
particles), and on their relative sign.

\section{Discussion}

\no We have studied the canonical quantization of the M.C.S.T. in a
path-representation. The physical sector of the theory, the basic gauge
invariant operators, and the Hamiltonian were explicitly calculated in this
geometric representation.  The resolution (eq.\ref{eq:soluliga}) of the Gauss
constraint (\ref{eq:ligacaminos}), provides a non-trivial example of path-space
calculation. Also, it shows the adventages of employing this formulation
to deal with  the geometrical content of the theory, which allows to relate
it with the quantum mechanics of point particles with anyonic behavior.                                                                                 

\no More preciselly, it is shown that the M.C.S.T. is equivalent  to
the theory of a massles scalar field  whose wave functional obeys anyonic
boundary conditions. This anyonic behavior is manifested in a simple form
within the path-representation framework, since the ends of the paths
(``points''
and ``anti-points'') just play the role of the particles whose exchanges 
give rise
to the non conventional statistical phase factor that reveals the anyonic
content of the theory. In other words, it is due to the fact that we are
working in a path-representation, instead of a ``shape-representation''
$|A_i\big >$, that we can make an easy contact with the model of anyonic
particles.

\no It would be interesting to explore the non-abelian counterpart of the
present theory in the corresponding geometric representation. One can
suspect that a non abelian ``signed-points'' representation, which arises
when dealing with the Principal Chiral Field, could be the key to carry out
this program (see comment after eq.(\ref{eq:unitaria})). It would also be
interesting to study the Self-Dual (i.e.: massive Chern-Simons)
theory \cite{townsend} in the path-representation. This model is dual (and
henceforth equivalent) to the M.C.S.T. \cite{stephany}, and probably there
exist an underlying geometry  supporting this duality that could be made
explicit with the aid of an appropriate geometric representation.

\vspace*{-9pt}

\end{document}